# Quantum state tomography with informationally complete POVMs generated in the time domain

Artur Czerwinski[1]

## Abstract

The article establishes a framework for dynamic generation of informationally complete POVMs in quantum state tomography. Assuming that the evolution of a quantum system is given by a dynamical map in the Kraus representation, one can switch to the Heisenberg picture and define the measurements in the time domain. Consequently, starting with an incomplete set of positive operators, one can obtain sufficient information for quantum state reconstruction by multiple measurements. The framework has been demonstrated on qubits and qutrits. For some types of dynamical maps, it suffices to initially have one measurement operator. The results demonstrate that quantum state tomography is feasible even with limited measurement potential.

**Keywords** Quantum state tomography · POVM · Random unitary dynamics · Quantum measurement

## 1 Introduction

Quantum communication and quantum computation require well-defined resources in order to encode quantum information [1–3]. For this reason, the ability to characterize quantum objects based on measurements is crucial in these fields [4,5]. Three types of quantum tomography are usually distinguished. Quantum state tomography aims at reconstructing the accurate state of a physical system [6] and then quantum process tomography with the goal of obtaining a description of evolution in terms of a dynamical map [7,8]. Finally, there is quantum measurement tomography which allows one to determine the characteristics of the actual operators governing the measurements [9]. In the present article, we focus on quantum state tomography, assuming that the dynamics of the system and measurement operators are well characterized.

✉ Artur Czerwinski
  aczerwin@umk.pl

1   Institute of Physics, Faculty of Physics, Astronomy and Informatics, Nicolaus Copernicus University, Grudziadzka 5, 87–100 Torun, Poland

Quantum state tomography originated in 1933 when Wolfgang Pauli posted the question whether a particle's wave function can be determined from the position and momentum probability densities [10]. However, the problem of recovering the state of a physical system dates back to 1852 when George G. Stokes introduced his famous four parameters which allow to uniquely determine the polarization state of a light beam [11]. For years there have been many proposals of tomographic techniques which differ in accuracy, types of measurement and other properties. The notion of quantum state tomography refers to reconstruction of any mathematical representation of a system, e.g., quantum wave functions, density matrices, state vectors, and Wigner functions. In this article, we focus on the tomographic reconstruction of an unknown density matrix [12,13].

According to fundamental postulates of quantum mechanics, any $d-$level physical system is associated with Hilbert space $\mathcal{H}$ such that $\dim \mathcal{H} = d$. Pure states of the system are represented by complex vectors from the Hilbert space, whereas mixed states are characterized by density operators, $\rho$, which belong to the space of all linear operators acting on $\mathcal{H}$ (denoted by $\mathcal{L}(\mathcal{H})$). By selecting a basis in the space, the density operators can be represented by a matrix which has to be Hermitian, positive semi-definite and trace one [14]. The set of all physical density matrices is called the state set and shall be denoted by $\mathcal{S}(\mathcal{H})$ (see more in [15]).

A standard approach to state tomography involves polarization measurements along with estimation methods such as *maximum likelihood estimation* (MLE) which guarantee that the output matrix belongs to the state set [16–19]. Another possibility to implement a feasible framework for state reconstruction is based on measurements defined by the mutually unbiased bases (MUBs) [20,21]. There are also tomographic techniques which solve the density matrix reconstruction problem by means of expectation values of Hermitian operators [22–24]. Finally, contemporary quantum state tomography methods usually utilize the concept of generalized quantum measurements, which is the focus of this article.

Regardless of the particular state reconstruction method one considers a total number of measurements as a resource. For this reason, economic frameworks, which aim at reducing the amount of required data, are gaining in popularity [25–28]. One of such approaches, i.e., the so-called stroboscopic tomography which inspired this article, utilizes information about quantum dynamics encoded in the generator of evolution, e.g., Ref. [29]. There have also been proposals involving continuous measurement defined in the time domain [30–33]. Special attention should be paid to the methods, both theoretical and experimental, which demonstrate the possibility to obtain the complete information about an unknown quantum state from a single measurement setup [34,35].

In this article, we contribute to the search for optimal methods of quantum tomography as we introduce a framework for dynamic generation of informationally complete set of measurement operators. In Sec. 2, we revise the definitions and concepts connected with the generalized quantum measurements. Then, in Sec. 3, we present the framework and in Sec. 4 we demonstrate its performance in state reconstruction of qubits subject to random unitary dynamics. Sec. 5 is devoted to qutrits tomography. The results prove that one can reduce the amount of resources needed for state tomography provided the system dynamics is known (given in the Kraus representation).

The framework has a potential to enhance the efficiency of experimental tomographic techniques.

## 2 Quantum measurement in state tomography

Properly defined measurements play a crucial role in quantum state tomography since they provide information about the unknown system. According to postulates of quantum mechanics, measurements are described by a collection $\{M_k\}$ of *measurement operators* [36]. The index $k$ refers to the results of measurement that may occur in the experiment.

In our framework, we postulate that the source can repeatedly perform the same procedure of preparing quantum systems in an unknown quantum state represented by a density matrix $\rho$. Thus, we have access to a relatively large number of identical quantum systems which are at our disposal. We can assume that each copy from the ensemble is measured only once. For this reason, the post-measurement state of the system is of little interest, whereas all attention is paid to the probabilities of the respective measurement outcomes. Therefore, we follow the *Positive Operator-Valued Measure* (POVM) formalism [37].

A general measurement is defined by a set of positive semi-definite operators $\{M_k\}$, acting on a finite-dimensional Hilbert space $\mathcal{H}$, such that

$$\sum_k M_k = \mathbb{I}, \tag{1}$$

where $\mathbb{I}$ denotes the identity operator. The set of operators $\{M_k\}$ is referred to as a POVM.

As already mentioned, in this approach we are not interested in the structure of post-measurement state. The set of operators $\{M_k\}$ is sufficient to determine the probabilities of the possible measurement outcomes which are computed according to Born's rule [38]:

$$p(k) = \text{Tr}\{M_k \rho\}. \tag{2}$$

Note that the probabilities have to sum up to one, i.e., $\sum_k p(k) = \sum_k \text{Tr}\{M_k \rho\} = 1$, which is equivalent to the condition Eq. 1 since $\text{Tr}\rho = 1$.

The probabilities Eq. 2 provide knowledge about an unknown state $\rho$. From a practical point of view, any experiment results in statistics of individual events. These counts are converted into relative frequencies by averaging them over time. Such experimentally extracted information corresponds to the mathematical probabilities which occur in theoretical frameworks. Therefore, it is justified to assume that the data of the form Eq. 2 can be derived from an experiment.

The goal of quantum state tomography is to estimate the state by using the results of measurement. If the measurement operators provide sufficient information for the state reconstruction, they are called a quorum [14]. In the case of a POVM, if the measurement provides complete knowledge about the state of the system, it is said to be an informationally complete POVM (IC-POVM) [39–41]. For a given system there might be various different sets of operators which lead to complete state characterization.

We shall follow an operational definition of an IC-POVM [42].

**Definition 1** (Informationally complete POVM) A POVM is called informationally complete (IC-POVM) if it consists of operators which span $\mathcal{L}(\mathcal{H})$.

As a consequence, an IC-POVM has to comprise at least $d^2$ operators. IC-POVMs which have exactly $d^2$ elements are called *minimal*. Note that a spanning set can be considered a generalized basis for a vector space: any vector from the space can be expressed as a linear combination of the set elements, but in general the set does not need to be composed of linearly independent (or normalized) vectors.

Usually, special attention is paid to a particular case of POVMs which is called a symmetric, informationally complete, positive operator-valued measure (SIC-POVM) [43]. Originally, SIC-POVMs are constructed from rank-one projectors, but their general properties have also been excessively studied [44].

**Definition 2** (Symmetric IC-POVM) Let us assume there is a set of $d^2$ normalized vectors $|\phi_k\rangle \in \mathcal{H}$ such that

$$|\langle \phi_j | \phi_k \rangle|^2 = \frac{1}{d+1} \ \text{ for } \ j \neq k. \tag{3}$$

Then, the set of rank-one projectors $\{\Pi_i\}$ defined as

$$\Pi_i := \frac{1}{d} |\phi_i\rangle\langle\phi_i| \ \text{ for } \ i = 1, \ldots, d^2 \tag{4}$$

constitutes a symmetric, informationally complete, positive operator-valued measure (SIC-POVM).

The concept of SIC-POVMs can be mathematically expressed in a very elegant manner by utilizing the frame theory [43]. Nonetheless, for the sake of the current analysis, the key property of SIC-POVMs relates to their maximum efficiency in quantum state estimation.

## 3 Framework for dynamic generation of IC-POVMs

Information stored in an unknown density matrix can be retrieved from a properly defined set of measurements. If one can implement measurements corresponding with an IC-POVM, then one obtains complete knowledge needed for density matrix reconstruction. Let us first assume that we do not have access to an IC-POVM or we want to perform quantum tomography by a fewer number of distinct measurement operators. Thus, at the beginning there is an **incomplete** set of positive semi-definite operators: $\mathcal{M} = \{M_1, M_2, \ldots, M_r\}$.

Starting from an incomplete POVM, we seek a way to dynamically generate an IC-POVM. We consider a scenario that there is an ensemble of identically prepared quantum systems which evolve and different operators can be measured at distinct time instants (each copy of the system is measured only once). In order to achieve this goal, we have to assume that we possess knowledge about the dynamics of the system. Changes in quantum systems are usually described by completely positive and trace-preserving (CPTP) maps, i.e., quantum channels. For the convenience of this analysis we shall adopt the Kraus representation [45,46].

**Definition 3** A linear map $\Lambda : \mathcal{L}(\mathcal{H}) \to \mathcal{L}(\mathcal{H})$ is a quantum channel if and only if:

$$\Lambda[X] = \sum_\alpha K_\alpha X K_\alpha^\dagger \quad \text{for all} \quad X \in \mathcal{L}(\mathcal{H})$$

and

$$\sum_\alpha K_\alpha^\dagger K_\alpha = \mathbb{I}_d. \tag{5}$$

The condition in Eq. 5 ensures that the CP map $\Lambda$ preserves trace. It is worth mentioning that the Kraus representation is non-unique.

In order to keep track of changes in the quantum system over time, one needs to introduce time-dependent CPTP maps $\Lambda_t$. Maps which are legitimate from the physical point of view are called *dynamical maps*.

**Definition 4** (Dynamical map) A continuous one-parameter family of maps $\{\Lambda_t, t \in R_+^1\}$ such that $\Lambda_t : \mathcal{L}(\mathcal{H}) \to \mathcal{L}(\mathcal{H})$ constitutes a dynamical map if and only if:

1. $\Lambda_t$ is completely positive for all $t \in R_+^1$,
2. $\Lambda_t$ is trace-preserving for all $t \in R_+^1$,
3. $\Lambda_0 = \mathbb{I}_d$,

where $R_+^1$ denotes the set of nonnegative real numbers.

The last condition in the definition 4 is natural, because if the family of maps $\{\Lambda_t, t \in R_+^1\}$ is to describe evolution of density operators, it has to satisfy the initial condition $\Lambda_0[\rho(0)] = \rho(0)$.

Bearing in mind the definition of the dynamical map, we shall assume that the evolution of our quantum system is given by a special kind a map, called *random unitary dynamics* [47–49].

**Definition 5** (Random unitary dynamics (RUD)) A dynamical map $\Lambda_t : \mathcal{L}(\mathcal{H}) \to \mathcal{L}(\mathcal{H})$ shall be called random unitary dynamics (RUD) if it has the form:

$$\Lambda_t[X] = \sum_\alpha \pi_\alpha(t) U_\alpha X U_\alpha^\dagger \tag{6}$$

where $U_\alpha$ are unitary matrices and $\{\pi_\alpha(t)\}$ is a time-continuous probability distribution, i.e.,

$$\sum_\alpha \pi_\alpha(t) = 1 \quad \text{and} \quad \pi_\alpha(t) \geq 0 \tag{7}$$

for all $t \geq 0$.

One should be aware that the terminology connected with the dynamical maps of the form Eq. 6 is not consistent. In [50], this type of dynamics is referred to as *random external fields*, whereas it can also be called *unital stochastic maps*, as in [51].

By applying the random unitary dynamics into the measurement result, we can write a time-dependent formula for the probability Eq. 2 associated with $k$th operator from the set $\mathcal{M}$:

$$p_k(t) = \text{Tr}\{M_k \rho(t)\} = \text{Tr}\left\{M_k \left(\sum_\alpha \pi_\alpha(t) U_\alpha \rho(0) U_\alpha^\dagger\right)\right\} = \\ = \sum_\alpha \pi_\alpha(t) \text{Tr}\{U_\alpha^\dagger M_k U_\alpha \rho(0)\}. \tag{8}$$

Note that the probability formula was rewritten due to algebraic properties of the matrix trace. Physically speaking, this is equivalent to the shift from the Schrödinger picture to the Heisenberg representation since the unitary operators $\{U_\alpha\}$ are now acting on $M_k$.

If the number of unitary operators $U_\alpha$ equals $\kappa$, we can select a discrete number of time instants $\{t_1, t_2, \ldots, t_\kappa\}$ and obtain a system of equations expressing probabilities:

$$p_k(t_1) = \sum_{\alpha=1}^{\kappa} \pi_\alpha(t_1) \text{Tr}\{U_\alpha^\dagger M_k U_\alpha \rho(0)\}, \\ p_k(t_2) = \sum_{\alpha=1}^{\kappa} \pi_\alpha(t_2) \text{Tr}\{U_\alpha^\dagger M_k U_\alpha \rho(0)\}, \\ \vdots \\ p_k(t_\kappa) = \sum_{\alpha=1}^{\kappa} \pi_\alpha(t_\kappa) \text{Tr}\{U_\alpha^\dagger M_k U_\alpha \rho(0)\}. \tag{9}$$

Let us formulate a theorem concerning the solvability of the system Eq. 9.

**Theorem 1** *From the system of equation Eq. 9 one can compute* $\{\text{Tr}\{U_\alpha^\dagger M_k U_\alpha \rho(0)\}\}$ *where* $\alpha = 1, \ldots, \kappa$ *if*

$$\det[\pi_\alpha(t_j)] \neq 0. \tag{10}$$

*Proof* One can observe that the system of equations Eq. 9 can be converted into a matrix equation with the matrix $[\pi_\alpha(t_j)]$ (where $\alpha = 1, \ldots, \kappa$ and $j = 1, \ldots, \kappa$) multiplying the vector of the unknown quantities. The condition $\det[\pi_\alpha(t_j)] \neq 0$ is sufficient for the matrix to be invertible, which leads to the ability to solve the system and, for each $k$, obtain the figures $\{\text{Tr}\{U_\alpha^\dagger M_k U_\alpha \rho(0)\}\}$ where $\alpha = 1, \ldots, \kappa$.

In general, the matrix $[\pi_\alpha(t_j)]$ does not have to be square. Nonetheless, in the context of quantum tomography, it does not seem sensible to consider a case with a number of measurements higher than $\kappa$. □

Due to repeated measurements of the same operator $M_k$ (over distinct copies of the system) we compute the set of figures $\{\text{Tr}\{U_\alpha^\dagger M_k U_\alpha \rho(0)\}\}$ where $\alpha = 1, \ldots, \kappa$.

Note that $U_\alpha^\dagger M_k U_\alpha$ is positive semi-definite since $M_k \geq 0$. Thus, the values $\{\text{Tr}\{U_\alpha^\dagger M_k U_\alpha \rho(0)\}\}$ can be considered as measured probabilities associated with the positive operators $\{U_\alpha^\dagger M_k U_\alpha\}$. In other words, repetition of one kind of measurement defined by the operator $M_k$ leads to a set of probabilities which are in accordance with the Born's rule and the POVM formalism.

The measurement procedure is performed for each operator from the initial set $\{M_1, M_2, \ldots, M_r\}$.

In order to be able to reconstruct the initial density matrix $\rho(0)$ from the probabilities $\text{Tr}\{U_\alpha^\dagger M_k U_\alpha \rho(0)\}$, the set of operators $\{U_\alpha^\dagger M_k U_\alpha\}$ where $k = 1, \ldots, r$ and $\alpha = 1, \ldots, \kappa$ has to span $\mathcal{L}(\mathcal{H})$, i.e., the space to which $\rho(0)$ belongs, and additionally:

$$\sum_{k=1}^{r} \sum_{\alpha=1}^{\kappa} U_\alpha^\dagger M_k U_\alpha = \mathbb{I}_d. \tag{11}$$

In other words, the set $\{U_\alpha^\dagger M_k U_\alpha\}$ where $k = 1, \ldots, r$ and $\alpha = 1, \ldots, \kappa$ has to constitute an informationally complete POVM. Note that some elements of the set $\{U_\alpha^\dagger M_k U_\alpha\}$ may be redundant. Thus, in specific circumstances one can either select a sufficient number of operators which shall constitute an IC-POVM or decide to implement an overcomplete set of measurement operators. The latter approach can be particularly facilitative in a realistic scenario when the measured probabilities are burdened with errors [18,52].

## 4 IC-POVMs for qubits subject to random unitary dynamics

### 4.1 Preliminaries

Let us consider a qubit whose evolution can be described by random unitary dynamics:

$$\rho(t) = \Lambda_t[\rho(0)] = \sum_{\alpha=0}^{3} \pi_\alpha(t) \sigma_\alpha \rho(0) \sigma_\alpha, \tag{12}$$

where $\sigma_0 = \mathbb{I}_2$ and $\{\sigma_1, \sigma_2, \sigma_3\}$ denotes the set of Pauli matrices. In the case of this dynamical map, the initial condition, i.e., $\Lambda_0 = \mathbb{I}_2$ implies that $\pi_0(0) = 1$.

The efficiency of the framework described in Sec. 3 depends on the analytical properties of the functions $\pi_\alpha(t)$. Thus, let us consider two specific examples which demonstrate the performance of the framework.

### 4.2 Example 1: qubit dephasing

Let us analyze a specific kind of qubit random unitary dynamics which is called dephasing. In this case we have:

$$\pi_0(t) = \frac{1 + e^{-\gamma t}}{2}, \quad \pi_1(t) = \pi_2(t) = 0, \quad \pi_3(t) = \frac{1 - e^{-\gamma t}}{2}, \tag{13}$$

which allows us to express the dynamical map as

$$\rho(t) = \frac{1+e^{-\gamma t}}{2}\rho(0) + \frac{1-e^{-\gamma t}}{2}\sigma_3\rho(0)\sigma_3 \quad (14)$$

where $\gamma > 0$ denotes a decoherence parameter.

For a quantum system with evolution given by Eq. 14 we can formulate and prove a theorem.

**Theorem 2** *For a qubit with dynamics given by Eq. 14 one can obtain an IC-POVM if probabilities of two positive operators $M_1$ and $M_2$ are measured at two time moments $t_1 \neq t_2$, where*

$$M_1 = \begin{bmatrix} \frac{1}{5} & \frac{1}{6} \\ \frac{1}{6} & \frac{1}{3} \end{bmatrix} \qquad M_2 = \begin{bmatrix} \frac{3}{10} & \frac{1}{7}+\frac{i}{10} \\ \frac{1}{7}-\frac{i}{10} & \frac{1}{6} \end{bmatrix}. \quad (15)$$

*Proof* First, one can quickly verify that $M_1, M_2 > 0$, which means that initially we have two positive operators at our disposal that do not constitute an IC-POVM. Thus, a single measurement of probability corresponding with each operator does not provide sufficient data for the density matrix reconstruction. Therefore, we see the need for the dynamic approach.

If the probability associated with the operator $M_1$ is measured at two time instants, we get a system of equations according to Eq. 9:

$$\begin{bmatrix} p_1(t_1) \\ p_1(t_2) \end{bmatrix} = \begin{bmatrix} \frac{1+e^{-\gamma t_1}}{2} & \frac{1-e^{-\gamma t_1}}{2} \\ \frac{1+e^{-\gamma t_1}}{2} & \frac{1-e^{-\gamma t_2}}{2} \end{bmatrix} \begin{bmatrix} \mathrm{Tr}\{M_1\rho(0)\} \\ \mathrm{Tr}\{(\sigma_3 M_1\sigma_3)\rho(0)\} \end{bmatrix}. \quad (16)$$

One can compute:

$$\det \begin{bmatrix} \frac{1+e^{-\gamma t_1}}{2} & \frac{1-e^{-\gamma t_1}}{2} \\ \frac{1+e^{-\gamma t_1}}{2} & \frac{1-e^{-\gamma t_2}}{2} \end{bmatrix} = \frac{e^{-\gamma t_1}}{2} - \frac{e^{-\gamma t_2}}{2} \neq 0, \quad (17)$$

which holds true based on the assumption that $t_1 \neq t_2$. This means that from the system Eq. 16 we can compute the figures $\mathrm{Tr}\{M_1\rho(0)\}$ and $\mathrm{Tr}\{(\sigma_3 M_1 \sigma_3)\rho(0)\}$ which, according to Born's rule Eq. 2, are equivalent to the measurement probabilities of the operators $M_1$ and $\sigma_3 M_1 \sigma_3$ for the unknown state $\rho(0)$.

In the same vein, one can write a matrix equation for double measurement of probability associated with the operator $M_2$, which can be solved under the same condition Eq. 17 and leads to figures: $\mathrm{Tr}\{M_2\rho(0)\}$ and $\mathrm{Tr}\{(\sigma_3 M_2 \sigma_3)\rho(0)\}$. Now one can observe that

$$\sigma_3 M_1 \sigma_3 = \begin{bmatrix} \frac{1}{5} & -\frac{1}{6} \\ -\frac{1}{6} & \frac{1}{3} \end{bmatrix} \equiv \tilde{M}_1$$

$$\sigma_3 M_2 \sigma_3 = \begin{bmatrix} \frac{3}{10} & -\frac{1}{7} - \frac{i}{10} \\ -\frac{1}{7} + \frac{i}{10} & \frac{1}{6} \end{bmatrix} \equiv \tilde{M}_2. \quad (18)$$

It is easy to verify that $M_1 + \tilde{M}_1 + M_2 + \tilde{M}_2 = \mathbb{I}_2$, which means that it remains to check whether the operators satisfy the spanning condition form Def. 1. One can consider the following equation in order to investigate whether the operators are linearly independent:

$$a\, M_1 + b\, \tilde{M}_1 + c\, M_2 + d\, \tilde{M}_2 = \begin{bmatrix} 0 & 0 \\ 0 & 0 \end{bmatrix}, \quad (19)$$

which can be equivalently expressed by means of a matrix equation:

$$\begin{bmatrix} \frac{1}{5} & \frac{1}{5} & \frac{3}{10} & \frac{3}{10} \\ \frac{1}{6} & -\frac{1}{6} & \frac{1}{7}+\frac{i}{10} & -\frac{1}{7}-\frac{i}{10} \\ \frac{1}{6} & -\frac{1}{6} & \frac{1}{7}-\frac{i}{10} & -\frac{1}{7}+\frac{i}{10} \\ \frac{1}{3} & \frac{1}{3} & \frac{1}{6} & \frac{1}{6} \end{bmatrix} \begin{bmatrix} a \\ b \\ c \\ d \end{bmatrix} = \begin{bmatrix} 0 \\ 0 \\ 0 \\ 0 \end{bmatrix}. \quad (20)$$

This matrix has a unique solution in the form $a = b = c = d = 0$, which implies that the operators $\{M_1, \tilde{M}_1, M_2, \tilde{M}_2\}$ are linearly independent. Since $\dim \mathcal{L}(\mathcal{H}) = 4$, then any set of four linearly independent operators spans the space $\mathcal{L}(\mathcal{H})$. This means that the set of operators $\{M_1, \tilde{M}_1, M_2, \tilde{M}_2\}$ is an IC-POVM, which finishes the proof ☐

As it was mentioned in Sec. 2, a given quantum state may be reconstructed from different IC-POVMs. Thus, the theorem 2 is not restricted to the proposed operators $M_1$ and $M_2$. One may expect that for a different pair of positive operators, we could be able to conduct a similar reasoning.

### 4.3 Example 2: general form of random unitary qubit evolution

Let us now consider a more general model of qubit dynamics such that the functions $\pi_\alpha(t)$ from Eq. 12 have the forms

$$\pi_0(t) = \frac{1 + e^{-\gamma_1 t} + e^{-\gamma_2 t} + e^{-\gamma_3 t}}{4}$$
$$\pi_1(t) = \frac{1 - e^{-\gamma_1 t}}{4}$$
$$\pi_2(t) = \frac{1 - e^{-\gamma_2 t}}{4}$$
$$\pi_3(t) = \frac{1 - e^{-\gamma_3 t}}{4},$$
(21)

where $\gamma_1, \gamma_2, \gamma_3$ are positive decoherence rates such that $\gamma_i \neq \gamma_j$ for $i \neq j$.

We propose to formulate and prove the following theorem.

**Theorem 3** *If qubits dynamics is given by the map Eq. 12 with $\{\pi_\alpha(t)\}$ defined in Eq. 21, then one can generate an IC-POVM by measuring the probability for the operator*

$$M_0 = \begin{bmatrix} \frac{1}{5} & \frac{1}{6} + \frac{i}{10} \\ \frac{1}{6} - \frac{i}{10} & \frac{3}{10} \end{bmatrix}$$
(22)

*at four time instants $t_1, t_2, t_3, t_4$ such that $t_i \neq t_j$ for $i \neq j$.*

***Proof*** First, one can notice that $M_0 > 0$, which means that this operator fits to the idea of generalized quantum measurement. Naturally, probability of one measurement $\text{Tr}\{M_0 \rho(0)\}$ would not be sufficient to determine an unknown density matrix $\rho(0)$. This justifies the need for the dynamic approach.

Let us assume that we are able to measure the probability of $M_0$ at four distinct time instants $\{t_1, t_2, t_3, t_4\}$, which gives us results denoted consecutively by $p(t_1), p(t_2), p(t_3), p(t_4)$. According to Eq. 9 the measurement scenario can be translated into a matrix equation:

$$\begin{bmatrix} p(t_1) \\ p(t_2) \\ p(t_3) \\ p(t_4) \end{bmatrix} = \begin{bmatrix} \Gamma_{11} & \Gamma_{12} & \Gamma_{13} & \Gamma_{14} \\ \Gamma_{21} & \Gamma_{22} & \Gamma_{23} & \Gamma_{24} \\ \Gamma_{31} & \Gamma_{32} & \Gamma_{33} & \Gamma_{34} \\ \Gamma_{41} & \Gamma_{42} & \Gamma_{43} & \Gamma_{44} \end{bmatrix} \begin{bmatrix} \text{Tr}\{M_0 \rho(0)\} \\ \text{Tr}\{(\sigma_1 M_0 \sigma_1)\rho(0)\} \\ \text{Tr}\{(\sigma_2 M_0 \sigma_2)\rho(0)\} \\ \text{Tr}\{(\sigma_3 M_0 \sigma_3)\rho(0)\} \end{bmatrix},$$
(23)

where $\Gamma_{i1} = (1 + e^{-\gamma_1 t_i} + e^{-\gamma_2 t_i} + e^{-\gamma_3 t_i})/4$ for $i = 1, 2, 3, 4$ and $\Gamma_{ij} = (1 - e^{-\gamma_{j-1} t_i})/4$ for $i = 1, 2, 3, 4$ and $j = 2, 3, 4$.

First, one can notice that since the functions $\pi_\alpha(t)$ are linearly independent (which can be demonstrated by calculating the Wronskian), the matrix $[\Gamma_{ij}]$ is invertible as long as $t_i \neq t_j$ for $i \neq j$. This means that experimentally accessible probabilities $p(t_1)$, $p(t_2)$, $p(t_3)$, $p(t_4)$ allow us to compute the set of figures: $\text{Tr}\{M_0 \rho(0)\}$, $\text{Tr}\{(\sigma_1 M_0 \sigma_1) \rho(0)\}$, $\text{Tr}\{(\sigma_2 M_0 \sigma_2) \rho(0)\}$, $\text{Tr}\{(\sigma_3 M_0 \sigma_3) \rho(0)\}$. Now one finds:

$$\sigma_1 M_0 \sigma_1 = \begin{bmatrix} \frac{3}{10} & \frac{1}{6} - \frac{i}{10} \\ \frac{1}{6} + \frac{i}{10} & \frac{1}{5} \end{bmatrix} \equiv \tilde{M}_1$$

$$\sigma_2 M_0 \sigma_2 = \begin{bmatrix} \frac{3}{10} & -\frac{1}{6} + \frac{i}{10} \\ -\frac{1}{6} - \frac{i}{10} & \frac{1}{5} \end{bmatrix} \equiv \tilde{M}_2 \quad (24)$$

$$\sigma_3 M_0 \sigma_3 = \begin{bmatrix} \frac{1}{5} & -\frac{1}{6} - \frac{i}{10} \\ -\frac{1}{6} + \frac{i}{10} & \frac{3}{10} \end{bmatrix} \equiv \tilde{M}_3,$$

from which it is easy to verify that $M_0 + \tilde{M}_1 + \tilde{M}_2 + \tilde{M}_3 = \mathbb{I}_2$. It remains to prove that the set of operators spans $\mathcal{L}(\mathcal{H})$. Let us investigate an equation: $a M_0 + b \tilde{M}_1 + c \tilde{M}_2 + d \tilde{M}_3 = \begin{bmatrix} 0 & 0 \\ 0 & 0 \end{bmatrix}$, which can be put into the matrix form:

$$\begin{bmatrix} \frac{1}{5} & \frac{3}{10} & \frac{3}{10} & \frac{1}{5} \\ \frac{1}{6} + \frac{i}{10} & \frac{1}{6} - \frac{i}{10} & -\frac{1}{6} + \frac{i}{10} & -\frac{1}{6} - \frac{i}{10} \\ \frac{1}{6} - \frac{i}{10} & \frac{1}{6} + \frac{i}{10} & -\frac{1}{6} - \frac{i}{10} & -\frac{1}{6} + \frac{i}{10} \\ \frac{3}{10} & \frac{1}{5} & \frac{1}{5} & \frac{3}{10} \end{bmatrix} \begin{bmatrix} a \\ b \\ c \\ d \end{bmatrix} = \begin{bmatrix} 0 \\ 0 \\ 0 \\ 0 \end{bmatrix}. \quad (25)$$

One can verify that the determinant of the left-hand side matrix is nonzero and, for this reason, the equation has one unique solution $a = b = c = d = 0$, which means that the operators $M_0$, $\tilde{M}_1$, $\tilde{M}_2$, $\tilde{M}_3$ are linearly independent and consequently they span $\mathcal{L}(\mathcal{H})$. Since they satisfy all necessary conditions, the operators constitute an IC-POVM and can be used as a source of information for quantum state tomography, which finishes the proof. □

### 4.4 Discussion and analysis

In the case of quantum tomography of qubits one traditionally needs to define four distinct operators in order to have an IC-POVM [36]. In practical applications, these four operators are often defined by means of polarization measurements [16,17,53]. In the current article, we have showed that, for qubits subject to dephasing, one can start with two positive operators and then dynamically generate an IC-POVM by double measurement of probabilities corresponding with these operators. This results suggest that we can reduce the amount of resource needed for quantum state tomography provided we can utilize the knowledge about system evolution.

The result concerning the second example proves that the effectiveness of the framework depends on the type of dynamics. For the qubit evolution given by the random unitary dynamics Eq. 12 with the probabilities $\{\pi_\alpha(t)\}$ defined in Eq. 21, one positive operator is sufficient to generate an IC-POVM. Such kind of dynamics can be considered *optimal* in the context of state recovery. Practically, it means that one needs to prepare one experimental setup and then repeat the same kind of measurement at four distinct time instants. Such a procedure provides complete knowledge about the unknown system. This approach appears to be more convenient than preparing a number of distinct experimental setups. This result is in line with the current discoveries which indicate that quantum state tomography based on one type of measurement is feasible [34,35].

It should also be noted that the positive operators proposed in the theorems are not the only ones that suit to the framework. IC-POVM can be realized by different sets of operators. Thus, one may expect that in the case of the theorem 3 the optimal state reconstruction would be possible if the initial operator $M_0$ was different than Eq. 22. A natural question may concern the possibility to implement an operator from the SIC-POVM in the dynamic framework. For dim $\mathcal{H} = 2$ the SIC-POVM is defined by four vectors (the set $\{|0\rangle, |1\rangle\}$ denotes the standard basis):

$$|\phi_1\rangle = |0\rangle \quad |\phi_2\rangle = \frac{1}{\sqrt{3}}|0\rangle + \sqrt{\frac{2}{3}}|1\rangle$$
$$|\phi_3\rangle = \frac{1}{\sqrt{3}}|0\rangle + \sqrt{\frac{2}{3}}e^{i\frac{2\pi}{3}}|1\rangle \quad |\phi_4\rangle = \frac{1}{\sqrt{3}}|0\rangle + \sqrt{\frac{2}{3}}e^{i\frac{4\pi}{3}}|1\rangle \tag{26}$$

and the operators are given by $\Pi_i = 1/2|\phi_i\rangle\langle\phi_i|$.

It can be proved that none of the operators $\Pi_1$, $\Pi_2$ can be substituted for $M_0$ from Eq. 22 since neither of the sets: $\{\Pi_i, \sigma_1\Pi_i\sigma_1, \sigma_2\Pi_i\sigma_2, \sigma_3\Pi_i\sigma_3\}$ for $i = 1, 2$ contains four linearly independent vectors. However, it turns out that either $\Pi_3$ or $\Pi_4$ can be implemented to generate an IC-POVM for a qubit subject to random unitary dynamics since we have:

$$\Pi_j + \sigma_1\Pi_j\sigma_1 + \sigma_2\Pi_j\sigma_2 + \sigma_3\Pi_j\sigma_3 = \mathbb{I}_2$$
$$\text{Span}\{\Pi_j, \sigma_1\Pi_j\sigma_1, \sigma_2\Pi_j\sigma_2, \sigma_3\Pi_j\sigma_3\} = \mathcal{L}(\mathcal{H}) \tag{27}$$

for both $j = 3, 4$. This observation means that when the measurement potential is limited one can perform only one kind of measurement, associated with either $\Pi_3$ or $\Pi_4$, at four distinct time instants in order to obtain complete information about the quantum state with dynamics defined by Eq. 21.

## 5 SIC-POVM for qutrits subject to random unitary dynamics

### 5.1 Preliminaries

Let us consider a qutrit subject to random unitary dynamics:

$$\rho(t) = \Lambda_t[\rho(0)] = \sum_{k=0}^{2}\sum_{l=0}^{2} \pi_{kl}(t) U_{kl} \rho(0) U_{kl}^\dagger, \tag{28}$$

where $\pi_{kl}(t)$ is defined:

$$\begin{aligned}\pi_{00}(t) &:= \frac{1 + \sum_{\alpha=1}^{8} e^{-\gamma_\alpha t}}{9} \quad \text{for} \quad (k, l) = (0, 0) \\ \pi_{kl}(t) &:= \frac{1 - e^{-\gamma_\alpha t}}{9} \quad \text{for} \quad \alpha \equiv (k, l) \neq (0, 0),\end{aligned} \tag{29}$$

and $\{\gamma_\alpha\}$ denotes 8 positive decoherence rates such that $\gamma_j \neq \gamma_i$ for $i \neq j$. In addition, $U_{kl}$ stands for 3−dimensional Weyl matrices:

$$\begin{aligned}
U_{00} &= \begin{bmatrix} 1 & 0 & 0 \\ 0 & 1 & 0 \\ 0 & 0 & 1 \end{bmatrix} & U_{01} &= \begin{bmatrix} 0 & 1 & 0 \\ 0 & 0 & 1 \\ 1 & 0 & 1 \end{bmatrix} & U_{02} &= \begin{bmatrix} 0 & 0 & 1 \\ 1 & 0 & 0 \\ 0 & 1 & 0 \end{bmatrix} \\
U_{10} &= \begin{bmatrix} 1 & 0 & 0 \\ 0 & \omega & 0 \\ 0 & 0 & \omega^2 \end{bmatrix} & U_{11} &= \begin{bmatrix} 0 & 1 & 0 \\ 0 & 0 & \omega \\ \omega^2 & 0 & 0 \end{bmatrix} & U_{12} &= \begin{bmatrix} 0 & 0 & 1 \\ \omega & 0 & 0 \\ 0 & \omega^2 & 0 \end{bmatrix} \\
U_{20} &= \begin{bmatrix} 1 & 0 & 0 \\ 0 & \omega^2 & 0 \\ 0 & 0 & \omega \end{bmatrix} & U_{21} &= \begin{bmatrix} 0 & 1 & 0 \\ 0 & 0 & \omega^2 \\ \omega & 0 & 0 \end{bmatrix} & U_{22} &= \begin{bmatrix} 0 & 0 & 1 \\ \omega^2 & 0 & 0 \\ 0 & \omega & 0 \end{bmatrix},
\end{aligned} \tag{30}$$

where $\omega = e^{2\pi i/3}$ and $\omega^2 = e^{-2\pi i/3} = \omega^\dagger$.

A common practice in quantum tomography of qutrits involves applying a SIC-POVM which for dim $\mathcal{H} = 3$ is characterized by means of the vectors [54]:

$$|\psi_0^0\rangle = \frac{1}{\sqrt{2}}(|0\rangle + |1\rangle), \quad |\psi_1^0\rangle = \frac{1}{\sqrt{2}}(e^{-2\pi i/3}|0\rangle + e^{2\pi i/3}|1\rangle),$$

$$|\psi_2^0\rangle = \frac{1}{\sqrt{2}}(e^{2\pi i/3}|0\rangle + e^{-2\pi i/3}|1\rangle),$$

$$|\psi_0^1\rangle = \frac{1}{\sqrt{2}}(|1\rangle + |2\rangle), \quad |\psi_1^1\rangle = \frac{1}{\sqrt{2}}(e^{-2\pi i/3}|1\rangle + e^{2\pi i/3}|2\rangle), \quad (31)$$

$$|\psi_2^1\rangle = \frac{1}{\sqrt{2}}(e^{2\pi i/3}|1\rangle + e^{-2\pi i/3}|2\rangle),$$

$$|\psi_0^2\rangle = \frac{1}{\sqrt{2}}(|0\rangle + |2\rangle), \quad |\psi_1^2\rangle = \frac{1}{\sqrt{2}}(e^{2\pi i/3}|0\rangle + e^{-2\pi i/3}|2\rangle),$$

$$|\psi_2^2\rangle = \frac{1}{\sqrt{2}}(e^{-2\pi i/3}|0\rangle + e^{2\pi i/3}|2\rangle),$$

where $\{|0\rangle, |1\rangle, |2\rangle\}$ denotes the standard basis in $\mathcal{H}$. These vectors allow one to define the measurement operators: $\Pi_i^j := 1/3|\psi_i^j\rangle\langle\psi_i^j|$ which satisfy all necessary conditions for a SIC-POVM. Thus, according to the standard approach if one wants to reconstruct the quantum state of a qutrit, one needs nine probabilities representing the outcomes of SIC-POVM.

### 5.2 Results: qutrits subject to random unitary dynamics

We can formulate and prove a theorem.

**Theorem 4** *Assuming that the evolution of a qutrit is given by the random unitary dynamics Eq. 28 with the functions $\pi_{kl}(t)$ defined as Eq. 29, one can extract complete knowledge about the initial state $\rho(0)$ from **one** operator $\Pi_i^j := 1/3|\psi_i^j\rangle\langle\psi_i^j|$ provided its probability is measured at nine distinct time instants.*

**Proof** Let us assume that we can prepare a measurement setup which gives us the probability associated with one elements from the SIC-POVM set: $\Pi_i^j := 1/3|\psi_i^j\rangle\langle\psi_i^j|$. The ensemble of quantum systems evolves and we perform nine measurements at distinct time instants: $\{t_1, \ldots, t_9\}$. The results of measurements, denoted by $\{p(t_1), \ldots, p(t_9)\}$, can mathematically be expressed by means of a matrix equation (cf. Eq. 23):

$$\begin{bmatrix} p(t_1) \\ p(t_2) \\ \vdots \\ p(t_9) \end{bmatrix} = \begin{bmatrix} \Gamma_{10} & \Gamma_{11} & \cdots & \Gamma_{18} \\ \Gamma_{20} & \Gamma_{21} & \cdots & \Gamma_{28} \\ \vdots & \vdots & \ddots & \vdots \\ \Gamma_{90} & \Gamma_{91} & \cdots & \Gamma_{98} \end{bmatrix} \begin{bmatrix} \mathrm{Tr}\{U_{00}^\dagger \Pi_i^j U_{00} \rho(0)\} \\ \mathrm{Tr}\{U_{01}^\dagger \Pi_i^j U_{01} \rho(0)\} \\ \vdots \\ \mathrm{Tr}\{U_{22}^\dagger \Pi_i^j U_{22} \rho(0)\} \end{bmatrix}, \quad (32)$$

where $\Gamma_{\beta\alpha} := \pi_\alpha(t_\beta)$. For the convenience of mathematical notation, the probability functions are labeled consecutively with one index: $\alpha = 0, \ldots, 8$.

First, one should notice that the matrix equation has a unique solution. Since the functions $\{\pi_\alpha(t)\}$ are defined as linearly independent, which can be demonstrated by calculating the Wronskian, we have $\det[\Gamma_{\beta\alpha}] \neq 0$ as long as $t_i \neq t_j$ for $i \neq j$. Thus, from the matrix equation Eq. 32 we can compute the figures $\text{Tr}\{U_{kl}^\dagger \Pi_i^j U_{kl}\, \rho(0)\}$, where $(i, j)$ is fixed and $k, l = 0, 1, 2$. The values computable from Eq. 32 can be understood as probabilities according to Born's rule since $U_{kl}^\dagger \Pi_i^j U_{kl} > 0$.

One can notice a connection between the Weyl matrices and the SIC-POVM operators:

$$U_{kl}^\dagger \Pi_i^j U_{kl} = \Pi_{i \oplus k}^{j \oplus l}, \tag{33}$$

where $\oplus$ denotes addition modulo 3. This property means that $U_{00}^\dagger \Pi_i^j U_{00} = \Pi_i^j$ and for any starting $\Pi_i^j$ the other operations let us generate the remaining eight operators from the SIC-POVM set. In other words, the set $\{U_{kl}^\dagger \Pi_i^j U_{kl}\}$ where $k, l = 0, 1, 2$ is equivalent to the SIC-POVM for any starting operator $\Pi_i^j$, which finishes the proof.
□

### 5.3 Discussion and analysis

The standard approach to quantum state tomography of qutrits requires the ability to realize nine distinct measurements associated with each operator from the SIC-POVM [55]. If one decides to reconstruct the quantum state of two entangled qutrits, then the number of measurement operators rises up to 81 [56]. In this section, we have proved that the number of distinct operators can be reduced if one implements the dynamic approach to quantum measurement.

The evolution model defined by Eq. 29 can be considered optimal since one measurement operator at our disposal is sufficient to generate the SIC-POVM. This can be achieved by the interdependence between the Weyl matrices and the SIC-POVM. Thus, we can start with only one element of the SIC-POVM and dynamically generate the remaining operators by multiple measurements. Although relations between the Weyl operators and SIC-POVMs have been studied [57], in this article we have proved that these algebraic properties can be applied to facilitate quantum state tomography.

## 6 Conclusions

We have proposed a dynamic approach to quantum measurement which can enhance the efficiency of quantum state tomography. The framework can reduce the amount of required resource since one can generate an IC-POVM (or SIC-POVM), starting from an incomplete set of measurement operators. Therefore, it offers a possibility to perform quantum state tomography with limited measurement potential. Alternatively, by generating additional measurements, the framework can facilitate both experimental state estimation, where there is a tendency to implement overcomplete sets of measurement operators [18,52], and numerical simulations in which a realistic measurement scenario is considered, e.g., Ref. [58].

The results correspond well to the current trends in quantum tomography which focus on reducing the number of distinct measurement setups needed for state recon-struction. A key difference between the present model and other proposals which rely on quantum dynamics is the fact that here we are investigating the quantum tomography problem in the time domain within the POVM formalism, whereas other works utilize either expectation values of Hermitian operators, cf. Ref. [22,29,31], or the concept of weak measurement, cf. Ref. [30,32]. Furthermore, the present framework is not restricted to specific types of density matrices (e.g., low rank), unlike other proposals which aim at minimizing the number of measurement operators, cf. Ref. [25,26].

Further research is needed to answer remaining questions concerning the model. First of all, more types of quantum dynamics should be tested in terms of their efficiency in generating IC-POVMs. Additionally, the framework shall be applied to multilevel quantum systems (e.g., entangled qubits and qutrits). Finally, the ultimate goal is to experimentally verify the efficiency of the framework.

**Acknowledgements** The author acknowledges financial support from the Foundation for Polish Science (FNP) (project First Team co-financed by the European Union under the European Regional Development Fund).

## Declarations

**Conflict of interest** The author declares that there is no conflict of interest.